\documentclass[a4paper,11pt]{article}
\usepackage{pos}
\usepackage{setspace}

\renewenvironment{thebibliography}[1]{
\begin{oldthebibliography}{#1}
  \setlength{\itemsep}{0em}
  \setlength{\parskip}{0em}
}
{
  \end{oldthebibliography}
}

\title{Axion-like particles and high-energy gamma rays: interconversion revisited}
\ShortTitle{ALPs and high-energy gamma rays}

\author*[a,b]{Rafael~{Alves Batista}}
\author[b]{Cristina Viviente}
\author[a,b]{Gaetano~{Di Marco}}
\author[a,b]{Miguel A. Sánchez-Conde}

\affiliation[a]{Instituto de Física Teórica UAM-CSIC, Universidad Autónoma de Madrid\\ C/ Nicolás Cabrera 13-15, 28049, Madrid, Spain}

\affiliation[b]{Departamento de Física Teórica, Universidad Autónoma de Madrid\\
M-15, 28049, Madrid, Spain}

\emailAdd{rafael.alvesbatista@uam.es}

\abstract{
Axion-like particles (ALPs) are hypothetical entities often invoked to solve various problems in particle physics to cosmology. They are one of the most promising candidates to explain the elusive dark matter. A way to search for ALPs is through their effects on photons. In the presence of external magnetic fields, ALPs and photons can convert into one another, leading to measurable signals. In this contribution we present results of Monte Carlo simulations of ALP-photon interconversion in magnetised environments. We focus on high-energy gamma rays with TeV energies travelling over cosmological distances. We include a full treatment of the intergalactic electromagnetic cascades triggered by the gamma rays. Finally, we discuss the impact of this improved treatment of the propagation for current and future ALP searches.

}

\ConferenceLogo{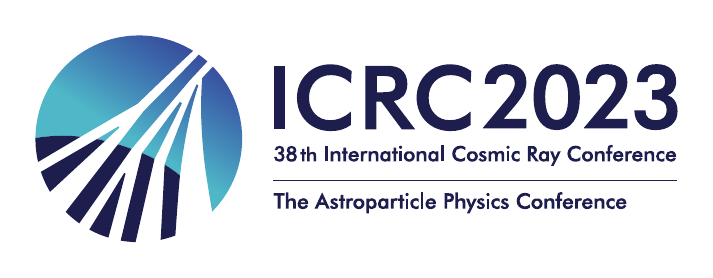}

\FullConference{%
The 38th International Cosmic Ray Conference (ICRC2023)\\
  26 July -- 3 August, 2023\\
  Nagoya, Japan}

\begin{document}
\maketitle

\section{Introduction}

The quest to find the elusive sources of dark matter has led to the exploration of various hypothetical particles beyond the Standard Model. Among these, axion-like particles (ALPs) have emerged as intriguing candidates, with far-reaching implications for both particle physics and cosmology. Initially proposed as extensions of the original axion concept~\cite{peccei1977a, peccei1977b}, which addressed the strong CP problem in quantum chromodynamics, these pseudo Nambu-Goldstone bosons have been postulated in various theoretical quantum gravitational frameworks~\cite{witten1984a, conlon2006a, arkanihamed2007a, arvanitaki2010a, cicoli2012a}.

ALPs have garnered significant attention as potential dark matter candidates. Despite their relatively low masses ($m_a \lesssim 1 \; \text{keV}$), if they are produced via some non-thermal mechanism in the primeval universe they could survive, forming the bulk of the cold dark matter believed to exist today~\cite{preskill1983a, sikivie1983a, dine1983a, sikivie2008a}.

A remarkable avenue for investigating ALPs lies in their interactions with photons during their propagation to Earth, which can leave discernible imprints in the spectra of gamma-ray sources~\cite{raffelt1988a}. Ordinarily, a very-high-energy (VHE) gamma-ray ($E \gtrsim 100 \; \text{GeV}$) from a cosmologically-distant source would be absorbed by pair-producing interactions ($\gamma + \gamma \rightarrow e^+ + e^-$) with pervasive background photon fields such as the extragalactic background light (EBL) or, at even higher energies ($E \sim 1 \; \text{PeV}$), the cosmic microwave background (CMB). However, in the presence of external magnetic fields, these photons would couple to  ALPs, depending on their mass ($m_a$) and coupling constant ($g_{a\gamma}$), following the coupling term present in the Lagrangian, of the form  $g_{a\gamma} \vec{E}_\gamma \cdot \vec{B} a$, wherein $\vec{E}_\gamma$ refers to the polarisation of the photon, $\vec{B}$ to the external magnetic field, and $a$ represents the ALP field. 

Signatures of photon-ALP mixing in gamma-ray data have been sought by several observatories, including the space-borne Fermi-LAT~\cite{fermi2016b} and imaging air-Cherenkov telescopes such H.E.S.S~\cite{hess2013c}. Furthermore, it is one of the goals of the Large High Altitude Air Shower Observatory (LHAASO)~\cite{bi2022a} and of several forthcoming projects like the Cherenkov Telescope Array (CTA)~\cite{cta2019a, cta2021b}, the ASTRI Mini-Array~\cite{vercellone2022a}, and the Southern Wide-field Gamma-ray Observatory (SWGO)~\cite{swgo2019a}.

In this contribution, we present the first results of a novel Monte Carlo simulation tool called \texttt{ALPinist}, which enables the modelling of photon-ALP mixing in various astrophysical environments. In section~\ref{sec:theory} we briefly describe the relevant theory of the mixing. Section~\ref{sec:alpinist} addresses the computational implementation of the code, followed by a few relevant examples in section~\ref{sec:simulations}. Finally, in sections~\ref{sec:discussion} and~\ref{sec:outlook} we discuss the perspectives for further developments and draw the conclusions (section~\ref{sec:outlook}).

\section{ALP--photon mixing theory}\label{sec:theory}

The mixing between ALPs and photons manifests through a coupling term in the Lagrangian
\begin{equation}
  \mathcal{L}_{a\gamma} = -\dfrac{1}{4} F_{\mu\nu} \tilde{F}^{\mu\nu} a = g_{a\gamma} \vec{E}_\gamma \cdot \vec{B} a \,
  \label{eq:lagrangian}
\end{equation}
where $a$ denotes the ALP field for a mass $m_a$, $g_{a\gamma}$ is a coupling constant, and $F_{\mu\nu}$ and $\tilde{F}^{\mu\nu}$ represents the electromagnetic tensor and its dual, respectively. 

The photon can be described by a linear combination of two polarisation modes, $A_1$ and $A_2$, which can be represented in a convenient basis, perpendicular to the propagation direction $\hat{z}$. Let us assume that $A_1$ essentially represents the direction of the intrinsic electric field of the photon, and that it forms an angle $\varphi$ with the component of the magnetic field component orthogonal to the direction of propagation.  In this case, the equation of motion for a photon in the presence of an ALP field ($a$) takes the form~\cite{raffelt1988a}
\begin{equation}
  \left( -E^2 - \dfrac{\partial^2}{\partial z^2} + \mathbb{M} \right) \vec{\mathcal{A}} = 0 \,,
  \label{eq:eom2}
\end{equation}
where $\vec{\mathcal{A}} \equiv (A_1, A_2, a)$, and the matrix $\mathbb{M}$ is given by
\begin{equation}
  \mathbb{M} \equiv 
  \begin{pmatrix}
    \Delta_\parallel \cos^2 \varphi +  \Delta_\perp \sin^2 \varphi & \left( \Delta_\parallel - \Delta_\perp \right) \sin\varphi \cos\varphi  & \Delta_{a\gamma} \cos\varphi \\
    \left( \Delta_\parallel - \Delta_\perp \right) \sin\varphi \cos\varphi &  \Delta_\parallel \sin^2 \varphi + \Delta_\perp \cos^2 \varphi & \Delta_{a\gamma} \sin\varphi \\
    \Delta_{a\gamma} \cos\varphi & \Delta_{a\gamma} \sin\varphi & \Delta_a \\
  \end{pmatrix} \,.
\end{equation}
The $\Delta$'s from this matrix for a traverse magnetic field of strength $B_t$ are defined as follows:
\begin{eqnarray}
  \Delta_{a} &=& - \left( \dfrac{1}{2} \dfrac{m_a c^2}{2 E} \right) ^ 2 \dfrac{1}{\hbar c} \,, \\
  \Delta_{a\gamma} &=& \dfrac{1}{2\mu_0} g_{a\gamma} B_t \sqrt{\hbar c} \,, \\
  \Delta_\perp &=& \Delta_\text{pl} + 4 \Delta_\text{QED} \,, \\
  \Delta_\parallel &=& \Delta_\text{pl} + 7 \Delta_\text{QED} \,.
\end{eqnarray}
The term $\Delta_\text{pl}$ is related to the plasma density of the medium:
\begin{equation}
  \Delta_\text{pl} = -\dfrac{n_e e^2}{2E} \dfrac{\hbar c \mu_0}{m_e} \,,
\end{equation}
where $m_e$ denoting the electron mass, $\mu_0$ is the vacuum permittivity constant, $e$ is the elementary charge of the positron, and $n_e$ refers to the electron density of the medium. The last quantity, $\Delta_\text{QED}$\footnote{Note that the convention adopted here is different from the one commonly found in the literature. We define $\Delta_\text{QED}$ as twice the values found in other works, to avoid rational coefficients for this term.}, arises from vacuum birefringence effects, but it tends to be subdominant for gamma-ray energies. 

Equation~\ref{eq:eom2} is a second-order differential equation. By noting that all relevant length scales are much larger than the wavelength of the gamma-ray photon, it can be recast into a first-order differential equation through the JWKB approximation (see ref.~\cite{raffelt1988a}), yielding
\begin{equation}
  \left( E - i \dfrac{\partial}{\partial z} + \dfrac{1}{2E}\mathbb{M} \right) \vec{\mathcal{A}} = 0 \,.
  \label{eq:eom}
\end{equation}
This is the equation of motion that will be used henceforth. 

For a simple homogeneous magnetic field, considering a photon propagating a distance $z$, the mixing resulting from solving eq.~\ref{eq:eom} is typically characterised by the so-called mixing angle ($\theta$):
\begin{equation}
  \theta = \dfrac{1}{2} \arcsin \left( \dfrac{2\Delta_{a\gamma}}{\Delta_\text{osc}} \right)\,,
\end{equation}
where
\begin{equation}
  \Delta_\text{osc} = \sqrt{\left(\Delta_a - \Delta_\text{pl}\right)^2 + 4\Delta_{a\gamma}^2} \,.
  \label{eq:deltaOsc}
\end{equation}
The probability of ALP-photon mixing, for this particular case, is then
\begin{equation}
  P_{a\gamma} = \sin^2 2\theta \sin^2 \left(\dfrac{1}{2} \Delta_\text{osc} z \right) \,.
  \label{eq:probOsc}
\end{equation}

Equation~\ref{eq:deltaOsc} can be thought as the wavenumber characterising the oscillatory behaviour of the photon-ALP interconversion at a given energy $E$. It can also be written in terms of a critical energy ($E_c$):
\begin{equation}
  \Delta_\text{osc} = 2 \Delta_{a\gamma} \sqrt{1 + \left(\dfrac{E_c}{E}\right)^2} \,,
  \label{eq:deltaOscEc}
\end{equation}
wherein
\begin{equation}
  E_c \equiv E \dfrac{\left| \Delta_a - \Delta_\text{pl} \right|}{2 \Delta_{a\gamma}} \,.
\end{equation}

Two regimes can be identified from eq.~\ref{eq:deltaOscEc}. For $E \gg E_c$, the so-called \emph{strong mixing regime}, $\Delta \simeq 2 \Delta_{a\gamma}$, such that $\theta \simeq 45^\circ$. However, at some energy, the QED vacuum polarisation term ($\Delta_\text{QED}$) starts to become relevant and dampens the oscillation. For $E \ll E_c$, there is essentially no mixing.

\section{The \texttt{ALPinist} code}\label{sec:alpinist}

\texttt{ALPinist} is a plugin for the CRPropa framework~\cite{alvesbatista2016a, alvesbatista2022a} that allows the treatment of ALP-photon mixing in various astrophysical environments. It takes advantage of the modular nature of CRPropa to facilitate the propagation of photons (or ALPs). The plugin is implemented in C++, with Python bindings, aligning with the underlying principles of CRPropa for seamless integration. Additionally, it offers built-in access to the extensive collection of magnetic-field models available in CRPropa, whilst also allowing for the utilisation of custom grids for both magnetic fields and plasma density.

To determine the state of the ALP-photon system at position $\vec{r}_j$, the fields $\vec{\mathcal{A}}$ at a point $\vec{r}_{j+1}$ can be calculated by solving equation~\ref{eq:eom}. For simplicity, let us assume that the propagation occurs along the $z$ direction, so that $\Delta z_j = \left| \vec{r}_{j+1} - \vec{r}_{j} \right|$. This equation can be easily solved by diagonalising the matrix $\mathbb{M}$, assuming that the magnetic field and plasma density remain approximately constant between $\vec{r}_{j}$ and $\vec{r}_{j+1}$.
From an algorithmic standpoint, this condition can be ensured by solving the equation of motion for infinitesimally small steps (labelled $j$), successively, until one of CRPropa's breaking conditions is met. These conditions may include the particle's energy dropping below a given threshold, traveling for an excessively long distance, or exiting a specific region.

For the propagation of gamma rays in intergalactic space, \texttt{ALPinist} can be combined with CRPropa modules responsible for modelling photon interactions and energy losses. One notable example is pair production resulting from interactions with pervasive photon fields.

\section{Simulations}\label{sec:simulations}

\begin{figure}[htb!]
  \centering
  \includegraphics[width=0.65\columnwidth]{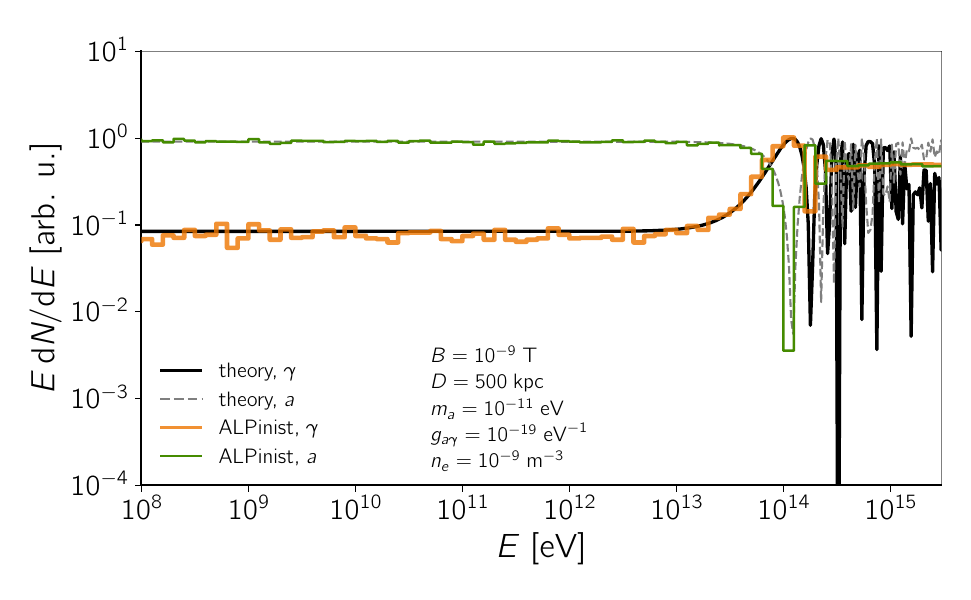}
  \caption{Histograms representing the results of the simulation of the propagation of 100,000 events (gamma rays)  over a distance of $500 \; \text{kpc}$ (orange and green lines). The parameters of the simulation are indicated in the figure. The black smooth lines represent the theoretical expectations for the flux of surviving photons (solid lines), and ALPs (dashed lines).}
  \label{fig:homogeneous}
\end{figure}

First, we present the results of a simulation of gamma-ray propagation in one dimension, over a distance $D = 500 \; \text{kpc}$. The magnetic field is assumed to be homogeneous, perpendicular to the direction of propagation, having a strength of $B = 1 \; \text{nT}$ ($=10 \; \mu\text{G}$). This simple scenario allows for a direct comparison with the theoretical predictions from eq.~\ref{eq:probOsc}, as shown in figure~\ref{fig:homogeneous}. 

Indeed, the agreement between simulation and theoretical expectations stemming from fig.~\ref{fig:homogeneous} is remarkable. Following the discussion from section~\ref{sec:theory}, we observe the onset of the strong mixing regime around $E \simeq 10 \; \text{TeV}$, followed by a prominent oscillatory behaviour.

\begin{figure}[hbt!]
  \centering
  \includegraphics[width=0.65\columnwidth]{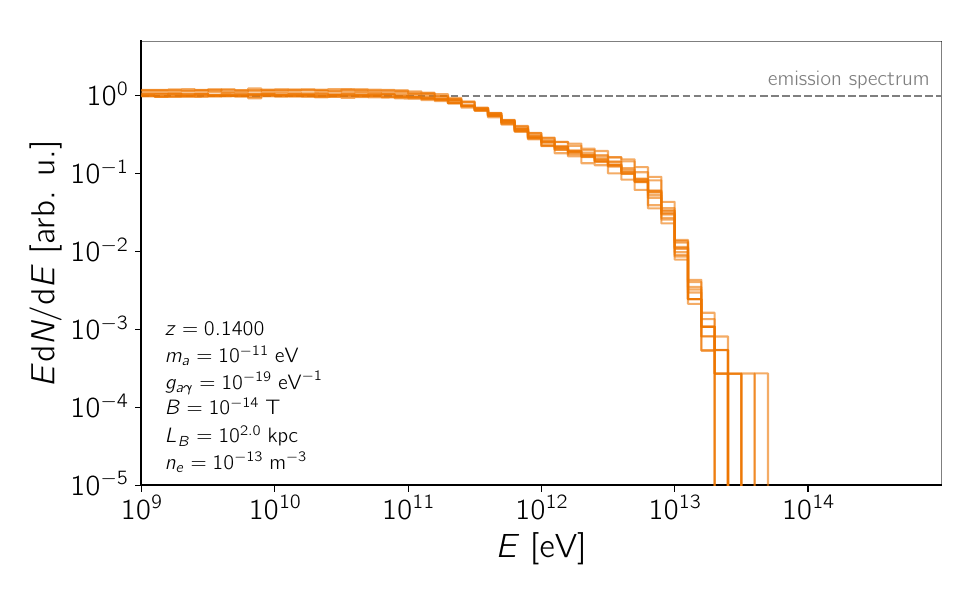}
  \caption{Histograms representing the results of the simulation of 500,000 events from an object at $z=0.14$ (solid orange lines). Each of the five lines correspond to an individual realisation of the magnetic field. The parameters of the simulation are indicated in the figure. The grey dashed line represent the intrinsic spectrum of the object.}
  \label{fig:turbulent}
\end{figure}

We now investigate a more realistic scenario. We consider an object at a redshift of $z=0.14$ emitting gamma rays. We consider pair production process with EBL and CMB photons since, at such large distances, a portion of the high-energy gamma-ray flux is directly impacted by these radiation fields. Furthermore, we consider a fairly strong ($B = 10^{-15} \; \text{T}$) Kolmogorov magnetic field with a coherence length of $L_B = 100 \; \text{kpc}$~\cite{alvesbatista2021a}. Once again, for simplicity, we focus on the one-dimensional case. Owing to the stochastic nature of the magnetic field, the results vary for different realisations. Therefore, in figure~\ref{fig:turbulent}, we present a family of curves representing each individual simulation.

\begin{figure}[htb!]
  \centering
  \includegraphics[width=0.495\columnwidth]{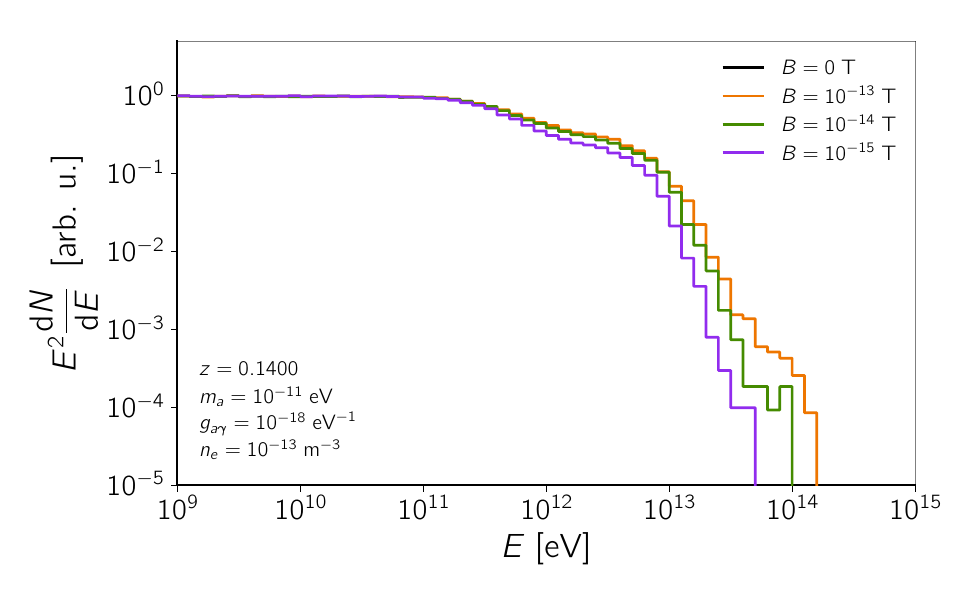}
  \includegraphics[width=0.495\columnwidth]{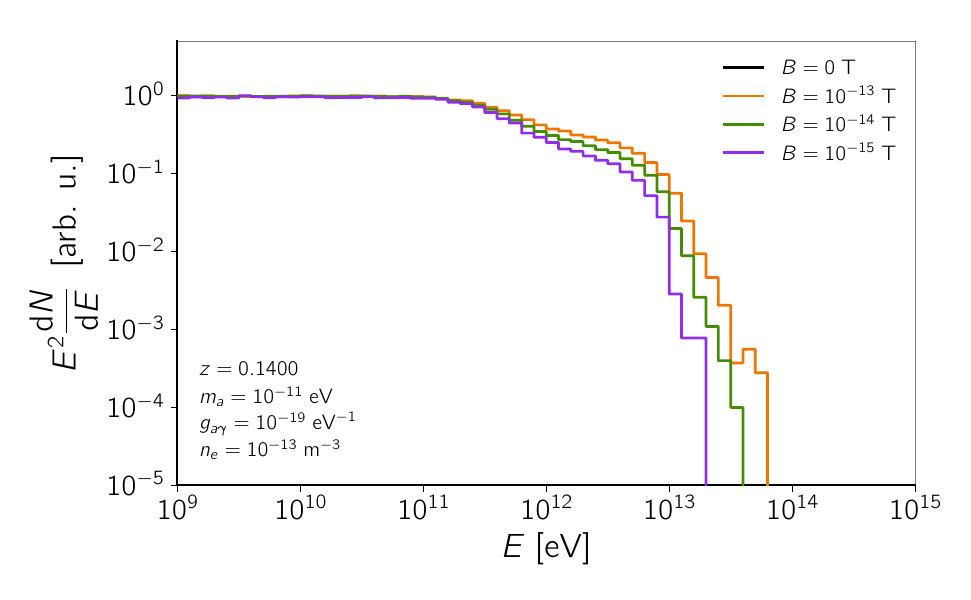}
  \caption{Expected photon flux from an astrophysical object at $z=0.14$ (solid orange lines) emitting according to a $E^{-2}$ power-law spectrum. Each line corresponds to the average of 20 magnetic field realisations. The left panel corresponds to the case of $g_{a\gamma} = 10^{-18} \; \text{eV}^{-1}$, whereas the right panel corresponds to $g_{a\gamma} = 10^{-19} \; \text{eV}^{-1}$, both for an ALP mass of $m_a = 10^{-11} \; \text{eV}$.
  The other parameters from the simulation are indicated in the figure.}
  \label{fig:simB}
\end{figure}

The scenario considered for the simulations shown in figure~\ref{fig:simB} is similar to that of figure~\ref{fig:turbulent}, also taking into account the pair production. We simulate the propagation of 100,000 events, and average the results of the simulation over 20 realisations of the magnetic field.

A remarkable feature of fig.~\ref{fig:simB} is the notorious suppression of the flux above some energy. This is due to the fact that gamma rays interact with the EBL through pair production. The effect of ALPs is evident: the universe would effectively become more transparent to the propagation of gamma rays, depending on the properties of the intervening magnetic fields, particularly for stronger ALP-photon coupling (left panel). 

\begin{figure}[hbt!]
  \centering
  \includegraphics[width=0.65\columnwidth]{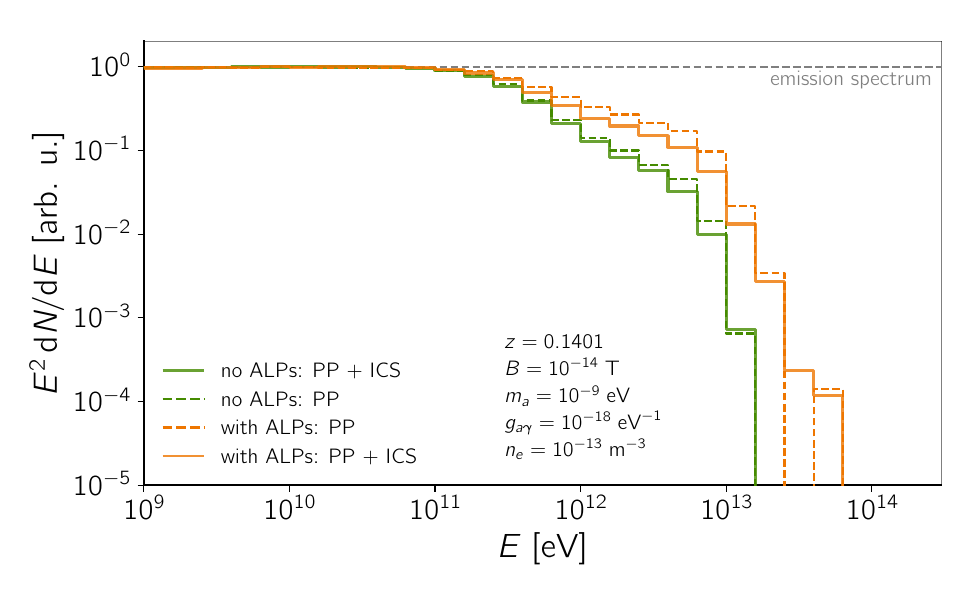}
  \caption{The expected photon flux from an astrophysical object at $z=0.14$, emitting according to an $E^{-2}$ power-law spectrum, average over 5 realisations of a magnetic field of strength $B=10^{-14} \; \text{T}$, is shown in the figure.  The ALP parameters from the simulation are indicated. The dashed lines represent simulations including only pair production, whereas the solid lines correspond to a combination of pair production and inverse Compton scattering. The orange lines include ALP-photon mixing, whereas the green lines do not.}
  \label{fig:simICS}
\end{figure}

We now generalise the results presented so far including, for the first time, inverse Compton scattering ($e^\pm + \gamma \rightarrow e^\pm + \gamma$) of the pair-produced electrons. To this end, we analyse one more scenario, as shown in figure~\ref{fig:simICS}. Note that the additional interaction results in a small but significant change in the expected gamma-ray fluxes compared to the case without inverse Compton scattering.

\section{Discussion}\label{sec:discussion}

In the examples presented in section~\ref{sec:simulations}, we have considered the propagation of photons, and hence the mixing, in a somewhat simplified fashion. A comprehensive model should take into account also the conversion due to magnetic fields within putative jets of the gamma-ray sources~\cite{mena2011a, davies2021a}, galaxy clusters~\cite{horns2012b, libanov2020a}, the Milky Way~\cite{troitsky2016a}, and in intergalactic space~\cite{mirizzi2007a}. The latter, in particular, is very poorly constrained~\cite{alvesbatista2021a}, such that its effect on the spectra of cosmological gamma-ray sources could be significant or non-existent. In reality, the magnetic fields in all of these regions ought to be taken into account when building reliable models~\cite{sanchezconde2009a, meyer2013a, meyer2014a}.

\texttt{ALPinist} is not the first Monte Carlo code for ALP-photon mixing. Extensions~\cite{kachelriess2022a} of the Elmag code~\cite{blytt2020a} also allow such studies. Other codes also exist, employing a semi-analytical approach~\cite{meyer2021a}. Nevertheless, \texttt{ALPinist} does have the advantage of enabling increasingly more complex treatments by calling \emph{all} relevant processes that affect photon propagation, for arbitrary energies, thanks to the modularity of CRPropa.

One drawback of the current implementation of \texttt{ALPinist} is that, being a Monte Carlo code, it is rather difficult to accurately capture oscillations at energies close to the critical energy with a fair sample of events. This can be seen in figure~\ref{fig:homogeneous}. At energies above $E \gtrsim 300 \; \text{TeV}$ an extremely fine binning would be necessary to reliably represent the predicted wiggles. At least from a simulation perspective, \texttt{ALPinist} allows the users to select the parameters to resolve these rapid oscillations.
While this level of detail is feasible for simulations focused solely on mixing and neglecting effects such as pair production, incorporating these effects would significantly increase simulation times, rendering them impractical. 

Another shortcoming of \texttt{ALPinist} is that, to include inverse Compton scattering, only the initially unpolarised cases can be simulated. However, an improved treatment of Compton interactions to address this limitation is currently underway.

\section{Conclusions and Outlook}\label{sec:outlook}

To enable \emph{realistic} modelling of ALP-photon mixing in various astrophysical environments, we have introduced the \texttt{ALPinist} code. This Monte Carlo code is a plugin for the CRPropa framework and enables detailed modelling of electromagnetic cascades in arbitrary astrophysical environments.
The main novelty of our work is the inclusion, for the first time, of the regeneration of photons due to inverse Compton scattering of pair-produced electrons and positrons. This is particularly important for studying ALP effects on the spatial morphology of gamma-ray haloes~\cite{alvesbatista2017d} around high-energy sources such as blazars.
Such precision is an important theoretical requirement~\cite{alvesbatista2021c} for interpreting upcoming observations~\cite{engel2022a} from instruments like the ASTRI  Mini-Array~\cite{vercellone2022a}, SWGO~\cite{swgo2019a}, and CTA~\cite{cta2019a}.

\footnotesize
\section*{Acknowledgements}
\noindent This work is funded by ``la Caixa'' Foundation (ID 100010434) and the European Union's Horizon~2020 research and innovation program under the Marie Skłodowska-Curie grant agreement No~847648, fellowship code LCF/BQ/PI21/11830030. It is also funded by grants PID2021-125331NB-I00
and CEX2020-001007-S, funded by MCIN/AEI/10.13039/501100011033, by ``ERDF A way of making Europe'', and the MULTIDARK Project RED2022-134411-T.


\end{document}